\documentclass[singlecolumn,journal]{IEEEtran}
\usepackage[T1]{fontenc}
\usepackage[misc]{ifsym}

\usepackage{mwe}
\usepackage[utf8]{inputenc}
\usepackage[colorlinks]{hyperref}
\usepackage{xcolor}
\usepackage{microtype}
\usepackage{amssymb}
\usepackage{amsmath}
\usepackage{mathtools}
\usepackage{ellipsis}
\usepackage{textcomp}
\usepackage{cite}
\usepackage{graphicx}
\usepackage[nolist,nohyperlinks]{acronym}
\usepackage{booktabs}
\usepackage{comment}
\usepackage{pifont}
\usepackage{adjustbox}
\usepackage{dsfont}
\usepackage{bm}
\usepackage{subcaption}
\usepackage{algorithm}
\usepackage[noend]{algpseudocode}
\usepackage{microtype}
\usepackage[capitalize,nameinlink]{cleveref}
\usepackage{tabularx}

\newcommand{\clip}{\operatorname{clip}}                     % clip(·)
\usepackage[normalem]{ulem}

\begin{document}

\title{Graph-Neural Multi-Agent Coordination for Distributed Access-Point Selection in Cell-Free Massive MIMO}

\author{
\IEEEauthorblockN{Mohammad Zangooei\IEEEauthorrefmark{1}, 
Lou Salaün\IEEEauthorrefmark{2}, 
Chung Shue Chen\IEEEauthorrefmark{2}, 
Raouf Boutaba\IEEEauthorrefmark{1}} \\
\IEEEauthorblockA{\IEEEauthorrefmark{1}University of Waterloo, Canada, 
Email: \{mzangooei, rboutaba\}@uwaterloo.ca} \\
\IEEEauthorblockA{\IEEEauthorrefmark{2}Nokia Bell Labs, Paris-Saclay, France, 
Email: \{lou.salaun, chung\_shue.chen\}@nokia-bell-labs.com}
}

\maketitle

\begin{abstract}
Cell-free massive MIMO (CFmMIMO) systems require scalable and reliable distributed coordination mechanisms to operate under stringent communication and latency constraints. A central challenge is the Access Point Selection (APS) problem, which seeks to determine the subset of serving Access Points (APs) for each User Equipment (UE) that can satisfy UEs' Spectral Efficiency (SE) requirements while minimizing network power consumption. We introduce APS-GNN, a scalable distributed multi-agent learning framework that decomposes APS into agents operating at the granularity of individual AP-UE connections. Agents coordinate via local observation exchange over a novel Graph Neural Network (GNN) architecture and share parameters to reuse their knowledge and experience. APS-GNN adopts a constrained reinforcement learning approach to provide agents with explicit observability of APS’ conflicting objectives, treating SE satisfaction as a cost and power reduction as a reward. Both signals are defined locally, facilitating effective credit assignment and scalable coordination in large networks. To further improve training stability and exploration efficiency, the policy is initialized via supervised imitation learning from a heuristic APS baseline. We develop a realistic CFmMIMO simulator and demonstrate that APS-GNN delivers the target SE while activating $50-70\%$ fewer APs than heuristic and centralized Multi-agent Reinforcement Learning (MARL) baselines in different evaluation scenarios. Moreover, APS-GNN achieves one to two orders of magnitude lower inference latency than centralized MARL approaches due to its fully parallel and distributed execution. These results establish APS-GNN as a practical and scalable solution for APS in large-scale CFmMIMO networks.
\end{abstract}


\section{Introduction}
Cell-free massive MIMO (CFmMIMO) has emerged as a compelling architecture for reliable wireless connectivity, in which many geographically distributed access points (APs) cooperatively serve all user equipments (UEs) without forming conventional cell boundaries~\cite{ngo2017cell}. By shifting from cell-centric to network-wide cooperation, CFmMIMO enables spatial diversity, mitigates inter-cell interference, and provides uniform coverage~\cite{demir2021foundations}. This cooperation is inherently challenging because APs are distributed; they must exchange information through limited fronthaul links and coordinate transmission decisions under tight coherence-time~\cite{freitas2024scalable, you2021distributed}.

Connecting a UE to a large set of APs improves link reliability, signal-to-interference-plus-noise ratio (SINR), and ultimately spectral efficiency (SE). However, it increases energy consumption, computational overhead, and fronthaul signaling. This trade-off naturally raises the following question: ``How should a CFmMIMO network determine, in a distributed manner, the subset of serving APs for each UE such that the UE’s SE requirement is satisfied while the overall power consumption is minimized?'' This problem is known as AP selection (APS) and poses a significant challenge~\cite{ngo2017total}. AP-UE connection decisions are tightly coupled through the wireless interference structure, leading to a combinatorial decision space where local decisions can have global performance implications.

A commonly used distributed method is the \texttt{k-Strongest} heuristic~\cite{buzzi2017cell}, where each UE connects to the $k$ APs with the strongest instantaneous channels~\cite{zaher2023soft, beerten2023cell}. While computationally simple and communication-efficient, this approach struggles under realistic conditions. Because channel strength fluctuates due to mobility, propagation randomness, and interference, the initially strongest APs may not remain so over time. In addition, a fixed $k$ ignores UE heterogeneity; some require more cooperative APs to satisfy SE targets, while others could be served by fewer. These limitations motivate APS mechanisms that explicitly adapt to channel dynamics and UE heterogeneity.

Reinforcement learning (RL) enables adaptive APS where the agent learns AP-UE connection activation strategies through repeated interaction with the wireless environment. However, formulating APS as a single-agent RL problem leads to an exponentially large action space and requires collecting global state information at a central controller, which is impractical in large deployments~\cite{ghiasi2022energy, mendoza2023user}. Single-agent RL also assumes a fixed network size, conflicting with the dynamic nature of real-world networks. These factors highlight the need for distributed RL approaches that scale gracefully with network size and respect communication and latency constraints.

We propose \textbf{APS-GNN}, a cooperative multi-agent reinforcement learning (MARL) framework that distributes the APS task across multiple agents, each responsible for activating a single AP-UE link. This design drastically reduces per-agent action spaces and enables experience reusing among agents for a scalable learning. More importantly, APS-GNN relies only on local observations and structured message passing among neighboring agents, allowing cooperation to emerge without global state aggregation. By leveraging graph neural networks (GNNs) to model the underlying AP-UE interaction topology, APS-GNN provides a communication-efficient mechanism for distributed coordination, making it suitable for large wireless systems. In addition to spatial coordination, APS-GNN accounts for temporal variability in wireless channels caused by UE mobility. Each agent processes a short history of its local channel magnitudes using a recurrent neural network (RNN) before engaging in GNN's message exchange.

APS-GNN adopts a \emph{constrained reinforcement learning} formulation rather than collapsing multiple objectives into a single scalar reward. In CFmMIMO systems, meeting SE requirements is a strict constraint, whereas reducing power consumption is an optimization objective. Combining these inherently different goals into a single reward requires extensive hand-tuning and lacks a principled mechanism for guaranteeing SE satisfaction. Instead, APS-GNN models SE violations as explicit costs and optimizes AP selection decisions subject to this constraint. To further improve credit assignment, we employ local reward and cost signals rather than globally aggregated values. APS-GNN is trained using a multi-agent Lagrangian Proximal Policy Optimization framework, in which dual critics estimate rewards and costs, and an adaptive Lagrange multiplier dynamically adjusts the policy’s attention to each of them. Finally, the policy parameters are shared across agents to reduce model complexity and improve learning efficiency.

To accelerate learning and improve stability, we pre-train APS-GNN using supervised learning before RL. Specifically, we construct a dataset based on a $k$-strongest heuristic, where each UE is connected to $k$ APs that provide the strongest channel and use these state-action pairs to train the actor following an imitation learning paradigm. This warm-start provides the policy with a meaningful baseline, reducing the inefficiency of random exploration and enabling RL to focus on balancing SE satisfaction with power efficiency.

To evaluate APS-GNN, we developed a CFmMIMO simulator that implements the system model described in this paper, including UE mobility, Line-of-Sight (LoS) channel computation, two-timescale APS and power control interactions, and interference modeling. Using this platform, we benchmarked APS-GNN against three representative families of AP selection strategies: (i) the $k$-Strongest heuristic (ii) PPO-Lagr, a centralized single-agent constrained RL formulation, and (iii) MAT-Lagr, a state-of-the-art transformer-based centralized MARL approach that we extended to support constrained RL.

Across all evaluated scenarios, APS-GNN reliably meets the target SE requirements while activating $50-70\%$ fewer APs than the best performing baseline. While APS-GNN achieves performance comparable to MAT-Lagr in medium-scale networks, it significantly outperforms MAT-Lagr in large-scale deployments, underscoring its superior scalability enabled by structured, local information sharing. In addition, execution-time measurements show that APS-GNN is one to two orders of magnitude faster than MAT-Lagr, making it well suited for real-world deployment. Ablation studies further demonstrate that constrained reinforcement learning, temporal modeling, and supervised pretraining each play a critical role in the observed performance gains. Collectively, these results highlight APS-GNN as a scalable, efficient, and practically deployable solution for AP selection in large-scale CFmMIMO networks.

\section{Related Work}
\textbf{Heuristic AP-UE Association.}
Early formulations of \mbox{CFmMIMO} assumed fully cooperative transmission, where all APs jointly serve every UE~\cite{ngo2017cell}. Although theoretically promising, such full connectivity is impractical because each AP must process and transport signals for all UEs, resulting in excessive fronthaul usage and power consumption. To improve scalability, Buzzi et al.~\cite{buzzi2017cell} proposed limiting cooperation so that each UE is served only by a subset of APs, giving rise to the AP-UE association problem. A widely used baseline is the $k$-Strongest heuristic~\cite{buzzi2017cell, zaher2023soft, beerten2023cell}, where each UE connects to the $k$ APs with the strongest instantaneous channel gains. While simple and communication-efficient, this method degrades under mobility, temporal fading, and heterogeneous channel conditions, since a fixed $k$ may over- or under-allocate APs across UEs. These drawbacks highlight the importance of adaptive APS strategies that can learn coordinated selection decisions based on local observations.

\textbf{Single-Agent RL and Centralized methods for AP Selection.}
Several works employ RL to improve energy efficiency and connection decisions. Ghiasi et al.~\cite{ghiasi2022energy} formulate APS as a single-agent RL problem to minimize power while meeting SE requirements, whereas Mendoza et al.~\cite{mendoza2023user} jointly optimize AP activation and AP-UE links using two single-agent RL formulations. Girycki et al.~\cite{girycki2023learning} apply RL to assign serving APs per resource block, aiming to maximize spectral efficiency. Ammar et al.~\cite{ammar2024handoffs} instead minimize UE handoffs using deterministic dynamic programming and suggest DRL for future extensions. Although effective in small networks, single-agent methods must operate on the full AP-UE association matrix, causing state and action spaces to scale quadratically with network size, limiting applicability in realistic deployments. Hao et al.~\cite{hao2024joint} propose a low-complexity, gradient-based method for jointly solving the power-allocation and AP-selection problems. Their solution is centralized and requires global network information, leading to substantial signalling overhead across the network. 

\textbf{Multi-Agent RL for AP Selection.}
Tsukamoto et al.~\cite{tsukamoto2023user, ikami2024distributed} propose a MARL-based AP-selection scheme in which each agent decides whether a UE should connect to any of its $k$ strongest APs. Their design, however, relies on a limited state representation that overlooks spatial correlations and temporal channel evolution, and the unconstrained RL setup requires manual reward tuning to satisfy SE or energy objectives. Similarly, Sun et al.~\cite{sun2023multi} use independent MARL agents to deactivate APs while aiming to maintain users’ SE, but without modeling system-level coordination. Banerjee et al.~\cite{banerjee2023access} also explore MARL for AP selection, where each AP runs an RL agent that decides which UEs to serve; however, agents do not communicate directly, and coordination is limited to a centralized critic. Although these methods improve scalability compared to single-agent RL, they lack structured information sharing, explicit constraint handling, and effective cooperation mechanisms.

\textbf{Supervised and Game-Theoretic Approaches.}
Pereira et al.~\cite{pereira2025gnn} propose a supervised GNN predictor that selects a subset of the $k$ strongest APs using a single message-passing layer. This method requires a labeled optimal AP selection dataset, which is expensive to generate at scale. Wei et al.~\cite{wei2025game} model APS as a stochastic exact potential game and develop DUAPSA, a solution where active UEs update AP-selection strategies using locally measured utilities and limited neighbor signaling. Although equilibrium existence is guaranteed, updates remain individually rational rather than globally cooperative and cannot seamlessly incorporate constraints or temporal dynamics.

\textbf{Contrast with APS-GNN.}
While prior work spans heuristic association algorithms, supervised learning, single-agent RL, independent MARL, and equilibrium-driven game-theoretic formulations, these approaches either lack coordination, do not scale to large networks, or depend on reward engineering and labeled data. In contrast, we formulate APS as a constrained multi-agent problem and propose \emph{APS-GNN}, a distributed framework that:
(i) assigns decision-making to AP-UE pairs, eliminating the state-action space explosion of single-agent methods;
(ii) enables structured cooperation through minimal GNN message passing, allowing agents to reason over spatial interference and network topology;
(iii) incorporates temporal channel dynamics via RNN-based state encoding; 
(iv) enforces SE and power objectives using constrained RL rather than manually tuned rewards, and
(v) accelerates training with a supervised heuristics-based pretraining.
Coupled with a realistic simulator that models mobility and two-timescale APS-power control interactions, APS-GNN provides a distributed, scalable, coordinated, and constraint-compliant solution for large CFmMIMO deployments.

\section{System Model and Problem Formulation} \label{sec:sys_model}

We consider a CFmMIMO system where \( M \) APs jointly serve \( K \) UEs in the downlink. The sets of AP and UE indices are defined as \( [M] \triangleq \{1, 2, \ldots, M\} \) and \( [K] \triangleq \{1, 2, \ldots, K\} \), respectively. All APs are connected to a centralized processing unit via fronthaul links, enabling coordinated transmission and signal processing. Each AP is equipped with a single antenna.

\subsection{Channel Model and Signal Transmission}
The channel between AP $m \in [M]$ and UE $k \in [K]$ at time step $t$ is characterized by a complex channel coefficient $g_{m,k}(t) \in \mathbb{C}$. The matrix of all channel coefficients is called \emph{channel matrix} and is denoted by $\bm{G}(t) \in \mathbb{C}^{M \times K}$.
\begin{equation}
\bm{G}(t) = 
    \begin{pmatrix}
    g_{1,1}(t) & \cdots & g_{1,K}(t) \\
    \vdots & \ddots & \vdots \\
    g_{M,1}(t) & \cdots & g_{M,K}(t)
    \end{pmatrix}
= \big( \bm{g}_1(t) \cdots \bm{g}_K(t) \big) 
\end{equation}

% \com{Talking about large-scaling fading here is confusing because all the formulas in this section use the instantaneous channel $\bm{G}$. Imo, this paragraph can be removed entirely}
% The wireless channel is modeled using large-scale fading effects, which capture long-term propagation characteristics such as path loss and shadowing. These components evolve slowly over space and time and are shaped by factors including transmitter-receiver distance, building density, antenna heights, and overall environmental morphology (urban, suburban, or rural). Their behavior is determined by system parameters such as carrier frequency and terrain geometry, and they provide the baseline attenuation profile used in our simulations \cite{series2009guidelines}.

We assume randomly distributed APs and UEs within a bounded area. The corresponding spatial coordinates are used to determine distances and angular relationships, which in turn influence the channel gains. In addition, user mobility is incorporated into the model by allowing user positions to evolve. This introduces temporal dynamics into the channel, reflecting how movement alters the propagation environment and thereby the channel coefficients. As users move, path loss and shadowing vary, leading to smooth temporal fluctuations in the large-scale fading.

% \com{I'm okay with reusing texts from our previous papers, but not too often to avoid being flagged by automatic plagiarism tools. So try to rephrase when possible.}
Let $\bm{q}(t) \in \mathbb{C}^K$ be the users’ message-bearing symbols to be transmitted. We assume that $\bm{q}(t)$ has a zero mean, unit variance and that the symbols are uncorrelated between users such that $\mathbb{E}(\bm{q} \bm{q^*}) = \bm{I}_K$, as commonly assumed in the literature. Firstly, $\bm{q}(t)$ is converted from the user data symbols space $\mathbb{C}^K$ to the precoded signals space $\mathbb{C}^M$ with $\bm{s}(t) = \bm{\Delta}(t) \bm{q}(t)$ where the precoding matrix is given as 
\begin{align} \label{eq:precoding}
    \; \bm{\Delta}(t) &= 
    \begin{pmatrix}
      \delta_{1,1}(t) & \cdots & \delta_{1,K}(t) \\
      \vdots & \ddots & \vdots \\
      \delta_{M,1}(t) & \cdots & \delta_{M,K}(t)
    \end{pmatrix} \notag \\
&= \left( \bm{\delta}_1(t) \cdots \bm{\delta}_K(t) \right) 
= 
    \begin{pmatrix}
    \overline{\bm{\delta}}_1^\mathsf{T}(t) \\ 
    \vdots \\
    \overline{\bm{\delta}}_M^\mathsf{T}(t)
    \end{pmatrix} \in \mathbb{C}^{M \times K},
\end{align}
and $\forall m \in [M]: \| \overline{\bm{\delta}}_m(t) \|_2 \leq 1$.

Let $\bm{x}(t) \in \mathbb{C}^K$ be the signal that users receive at time step $t$. It can be modeled as

\begin{equation} \label{eq:singal_propagation}
    \bm{x}(t) = \bm{G}^\mathsf{T}(t) \sqrt{\rho_d} \bm{s}(t) + \bm{w}(t),
\end{equation}
where $\rho_d$ is the maximum transmit power for each AP divided by the noise power $N_0$, $\bm{s}(t) \in \mathbb{C}^M$ is the power normalized precoded signal to be transmitted by the $M$ APs and $\bm{w}(t) \in \mathbb{C}^K$ is a circularly symmetric Gaussian noise vector. From Equations \eqref{eq:precoding} and \eqref{eq:singal_propagation}, we get $\bm{x}(t) = \sqrt{\rho_d} \bm{G}^\mathsf{T}(t) \bm{\Delta}(t) \bm{q}(t) + \bm{w}(t)$, which gives the received signal at the $k$-th user as 
\begin{align}
&x_k(t) = \sqrt{\rho_d} \bm{g}_k^\mathsf{T}(t) \bm{\Delta}(t) \bm{q}(t) + w_k(t) \\ 
&= \sqrt{\rho_d} \bm{g}_k^\mathsf{T}(t) \bm{\delta_k}(t) q_k(t) + \sqrt{\rho_d} \sum_{l \neq k} \bm{g}_k^\mathsf{T}(t) \bm{\delta_l}(t) q_l(t) + w_k(t). \nonumber
\end{align}

The first term represents the useful signal intended for user $k$, while the second term aggregates multi-user interference originating from the signals intended for all $l \neq k$. The final term accounts for thermal noise. This decomposition captures a central challenge in CFmMIMO: each user receives not only its intended beamformed signal but also residual leakage from beams serving other users. To quantify link quality, we use the signal-to-interference-plus-noise ratio (SINR). Formally, the SINR of UE $k$ at time $t$ is
\begin{equation}
    \mathrm{SINR}_k(t) = \frac{\rho_d |\bm{g}_k^\mathsf{T}(t) \bm{\delta}_k(t)|^2}{1 + \rho_d \sum_{l \neq k} |\bm{g}_k^\mathsf{T}(t) \bm{\delta}_l(t)|^2} \cdot
\end{equation}

Finally, the spectral efficiency (SE) translates the physical-layer SINR into the data rate achievable per unit bandwidth. Under the Shannon capacity approximation, the SE of user $k$ is expressed as
\begin{equation}
    \mathrm{SE}_k(t) = \log_{2}\!\left(1 + \mathrm{SINR}_k(t)\right) \quad \text{[bits/s/Hz]}.
\end{equation}

Higher SINR values yield higher SE, reflecting the fundamental tradeoff between interference management, channel quality, and throughput.

\subsection{Precoding, Power Control and AP Selection} 
In practice, precoding and AP-UE association operate on different timescales because they respond to different forms of channel variability. Precoding adapts to small-scale fading and instantaneous interference levels, both of which fluctuate rapidly over time. As a result, updating $\bm{\Delta}(t)$ at every time step enables the system to track these fast variations and maintain reliable per-slot performance.

By contrast, selecting the subset of APs that serve each UE depends primarily on large-scale channel characteristics such as path loss, shadowing, and user position. These parameters evolve slowly relative to the slot duration. Recomputing AP-UE associations too frequently would incur unnecessary signaling and coordination overhead, while providing negligible performance gains. Therefore, AP assignments are updated on a slower timescale, typically once every $T$ time steps, where $T$ is chosen to match the coherence time of the large-scale fading.

Mathematically, we define the AP selection matrix as $\bm{A}(t)$ with binary variables $\alpha_{m,k}(t)$ capturing the connection status of AP $m$ and UE $k$ at time step $t$. The AP selection is conducted at $t=0, \;T, \;2T, \;\cdots$ and held fixed in the intervening slots:
\begin{equation}
    \bm{A}(t) = \bm{A}(\lfloor t / T \rfloor)
\end{equation}

This structure constrains power control by ensuring that an AP allocates power only to the UEs it is assigned to serve:
\begin{equation}
    \alpha_{m,k}(t) = 0 \implies \delta_{m,k}(t) = 0
\end{equation}

Within this framework, we adopt Maximum Ratio Transmission (MRT) for downlink power control and precoding. Under MRT, each AP scales the signal for user $k$ by the conjugate of its locally estimated channel, requiring only local channel information obtained from uplink pilots and avoiding any inter-AP communication. Each AP allocates equal power among the UEs it serves, ensuring a simple and balanced transmission strategy. Owing to its simplicity, low computational cost, and fully distributed nature, MRT is well suited to large-scale cell-free deployments, allowing each AP to compute its transmit weights independently while still achieving strong array gain.

We consider a practical energy consumption model that considers electromagnetic radiation and hardware-related fixed power~\cite{ngo2017total}. Total power consumption is the summation of power consumption at each AP: $P_{\textit{tot}}(t) = \sum_{m=1}^{M} P_{m}(t)$.
The power consumption at the APs consists of (a) circuit power $P_{\textit{ap}}$ that captures RF chains and amplifiers' power consumption, and (b) signal transmission power. If AP $m$ does not serve any user at time step $t$, it will shut down to save unnecessary hardware energy consumption: $P_{\textit{ap}, m}(t) = P_{\textit{ap}} \times \operatorname{sign}{\!\left(\sum_{k=1}^{K}\alpha_{m,k}(t)\right)}$. Therefore, the power consumed by AP $m$ at time step $t$ can be written as follows
\begin{equation} \label{eq:power_consumption}
    P_{m}(t) = P_{\textit{ap}} \times \operatorname{sign}{\!\left(\sum_{k=1}^{K}\alpha_{m,k}(t)\right)} + \frac{\rho_d N_0}{\gamma_m}\sum_{k=1}^{K} |\delta_{m,k}(t)|^2,
\end{equation}
where $\rho_d N_0$ is the AP's maximum transmit power and $0 < \gamma_m < 1$ represents its power amplifier efficiency. According to MRT, each AP transmits at full power, therefore $\sum_{k=1}^{K} |\delta_{m,k}(t)|^2 = 1, \; \forall m \in [M]$ in the above formula.

\subsection{Problem Formulation}

We study an \emph{online} APS problem in a CFmMIMO network, where the objective is to minimize power consumption while ensuring that every UE meets a homogeneous SE target~$\kappa$ in expectation. Time is divided into consecutive windows of length $T$, and the AP configuration for window~$n$ is chosen once at the beginning of that window. Within each window, the selected configuration remains fixed, while channel conditions fluctuate due to UE mobility and other random propagation effects.

At the beginning of window~$n$, denoted by $[T_n] = [nT,(n+1)T-1]$), we determine the binary AP selection matrix $\bm{A}^{(n)}$ by solving:
\begin{subequations}\label{eq:opt_problem}
\begin{align}
\min_{\bm{A}^{(n)}}\quad
& \mathbb{E}\!\left[\sum_{t\in[T_n]} \sum_{m=1}^{M} P_m(t)\right] \label{eq:obj} \\
\text{s.t.}\quad & \mathbb{E}\!\left[\frac{1}{T} \sum_{t\in[T_n]} \mathrm{SE}_k(t)\right] \ge \kappa, && \forall k \in [K], \label{eq:se_req} \\
& \bm{A}(t)\,\bm{\Delta}(t) = \bm{\Delta}(t), && \forall t\in[T_n], \label{eq:ap_enforcement} \\
& \bm{A}(t) = \bm{A}^{(n)}, && \forall t\in[T_n]. \label{eq:two_timescale}
\end{align}
\end{subequations}

Constraint~\eqref{eq:se_req} ensures that each UE receives its required average SE over the window. Constraint~\eqref{eq:ap_enforcement} restricts APs to serving only the UEs they are assigned to in $\bm{A}^{(n)}$. The two-timescale condition~\eqref{eq:two_timescale} captures the fact that AP selection is updated once per window, whereas power allocation and precoding adapt at each slot.

% Under MRT power control, any active AP distributes its power among the UEs it serves, whereas a deactive AP transmits no power. Thus, activating an AP effectively incurs its full power budget, making the objective in~\eqref{eq:obj} closely aligned with selecting the smallest set of APs that can satisfy the SE constraints in expectation.

Under MRT power control, each AP allocates its transmit power among the UEs it is selected to serve, while an AP that is not selected by any UE consumes no power. Consequently, AP activation is not an explicit control variable but an implicit outcome of the per UE-AP selection decisions. An AP incurs its full power budget as soon as it serves at least one UE. As such, the objective in~\eqref{eq:obj} aligns with identifying the smallest subset of APs that are sufficient for satisfying the SE constraints in expectation.

Table~\ref{table:notation} summarizes the key notation used throughout the system model and problem formulation for clarity and reference.

\begin{table}[!h]
    \caption{Summary of notation used in system model and problem formulation}
    \centering
    \begin{tabularx}{0.47\textwidth}[t]{@{}l@{\quad}X@{}}
        \toprule
        \textbf{$M$} & Number of APs in the network \\
        \textbf{$K$} & Number of UEs in the network \\
        \textbf{$[M], [K]$} & AP and UE index sets, respectively \\
        \textbf{$g_{m,k}(t)$} & Complex channel between AP $m$ and UE $k$ at time $t$ \\
        \textbf{$\bm{G}(t)$} & Channel matrix $\in \mathbb{C}^{M \times K}$ at time $t$ \\
        \textbf{$\bm{q}(t)$} & Vector of user data symbols $\in \mathbb{C}^{K}$ \\
        \textbf{$\bm{\Delta}(t)$} & Precoding matrix $\in \mathbb{C}^{M \times K}$ \\
        \textbf{$\bm{s}(t)$} & Transmit signal vector at the APs \\
        \textbf{$\bm{x}(t)$} & Received signal vector at the UEs \\
        \textbf{$\bm{w}(t)$} & Additive Gaussian noise vector at time $t$ \\
        \textbf{$\rho_d$} & Maximum AP transmit SNR (power over noise) \\
        \textbf{$\mathrm{SINR}_k(t)$} & SINR of UE $k$ at time $t$ \\
        \textbf{$\mathrm{SE}_k(t)$} & Spectral efficiency of UE $k$ at time $t$ \\
        \textbf{$\alpha_{m,k}(t)$} & Binary AP-UE association variable \\
        \textbf{$\bm{A}(t)$} & AP selection matrix at time $t$ \\
        \textbf{$T$} & AP-selection window length (time steps) \\
        \textbf{$[T_n]$} & Time-slot set for window $n$ \\
        \textbf{$P_m(t)$} & Total power consumption of AP $m$ at time $t$ \\
        \textbf{$P_{\textit{tot}}(t)$} & Network-wide power consumption at time $t$ \\
        \textbf{$P_{\textit{ap}}$} & Fixed circuit power per active AP \\
        \textbf{$\gamma_m$} & Power amplifier efficiency of AP $m$ \\
        \textbf{$\kappa$} & Target average SE requirement per UE \\
        \bottomrule
    \end{tabularx}
    \label{table:notation}
\end{table}

\section{Multi-agent Design of APS-GNN} \label{sec:design}

The APS problem in \eqref{eq:opt_problem} is NP-hard~\cite{ghiasi2022energy} because of the binary AP-UE connection status variables that grow rapidly with the number of APs and UEs and their coupling with SE and power objectives. The problem complexity is exacerbated by time-varying channel conditions as UE positions, path loss, and interference fluctuate. 

To address the APS problem, we propose APS-GNN, a cooperative MARL framework that leverages GNNs for structured information exchange among agents~\cite{sutton2018reinforcement, gronauer2022multi}. APS-GNN enables distributed access-point decision making while capturing the underlying graph-structured dependencies induced by large-scale CFmMIMO networks. In the following, we present the architecture, learning formulation, and training procedure of APS-GNN in detail.

\subsection{Rationale for Multi-Agent Design}
APS-GNN assigns one agent to each AP-UE pair. Each agent makes a binary decision $a \in \{0,1\}$, resulting in $M \times K$ agents acting over low-dimensional action spaces. This is in contrast to a single-agent RL formulation that treats the entire association matrix $\bm{A} \in \{0,1\}^{M \times K}$ as one joint action. This decomposition provides several benefits that we discuss below.

First, it enables experience sharing across agents that solve structurally identical subproblems. This way the experience collected in one agent, or sub-problem, can be immediately utilized by others. This substantially improves sample efficiency since all agents use each other's experiences. 

Second, such a design eliminates redundancy due to permutation symmetry. In a single-agent formulation, the agent must implicitly learn that different permutations of the same AP-UE correspond to equivalent states or decisions, leading to unnecessary exploration of symmetric but behaviorally identical configurations. A multi-agent formulation collapses these equivalent permutations into a single representation, thereby reducing the effective exploration space and accelerating convergence. 

Third, the approach significantly reduces the number of trainable parameters. Rather than learning a monolithic policy that scales with the size of the system, the proposed design learns a single local policy that is reused across agents. Moreover, this structure naturally accommodates the addition or removal of APs and UEs without policy retraining, making the approach resilient to topology changes and mobility.

Fourth, restricting each agent’s observations to information relevant to its own AP-UE link and its local neighborhood reflects the inherent spatial locality of wireless network operations. Distant APs and UEs typically have negligible impact on a given link’s performance and therefore need not be considered. Limiting information exchange to local neighborhoods reduces observation noise and improves the agent’s ability to extract meaningful patterns. In Section~\ref{subsec:gnn}, we show how this localized information sharing can be efficiently realized using GNN-based structured state representations.

Lastly, the multi-agent design enables a decentralized operation where each agent decides based on the local information. A centralized controller must gather global channel information leading to prohibitive signaling overhead and increased latency. By contrast, distributing the decision process across many agents reduces reliance on network-wide signaling and scales gracefully with the network size. 

\subsection{Channel Information Processing} \label{subsec:gnn}
\begin{figure}[t] 
\includegraphics[width=0.9\linewidth]{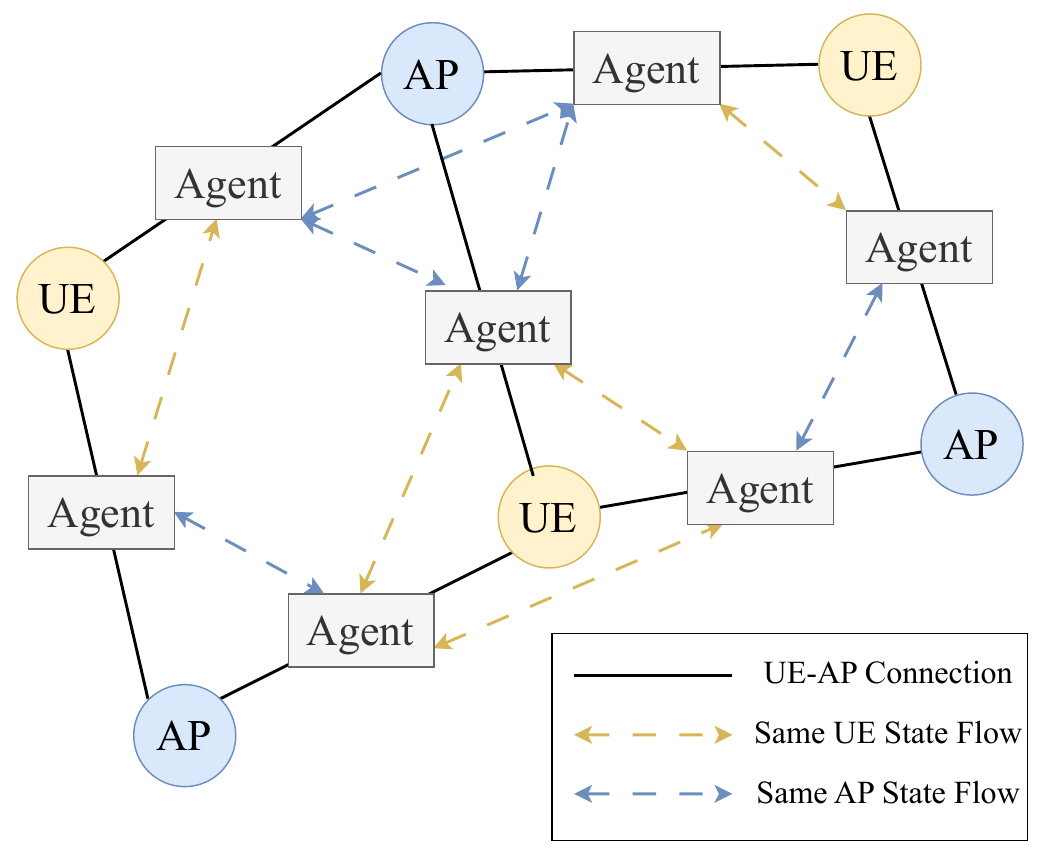}
\caption{Message-passing structure among agents: yellow dashed arrows for agents that share the same UE states and blue dashed arrows for agents that share the same AP.}
\label{fig:state}
\end{figure}

Each agent’s local observation includes the channel magnitude between its AP and UE. To promote coordinated behavior among agents, we leverage two structured interaction types:
(i) agents serving the same UE exchange information to collectively meet the UE’s SE target; and
(ii) agents sharing the same AP communicate to coordinate the activation process and manage power expenditure.
These interactions form a structured communication graph among agents, as illustrated in Figure~\ref{fig:state}. This graph structure has been shown to efficiently process channel information for several tasks in CFmMIMO~\cite{parlier2024learning, salaun2022gnn, mishra2024graph}.

We implement the message passing over two types of edges in a GNN. Each message is weighted by a single-head softmax attention coefficient. Then, the messages are aggregated at the destination node using the average followed by a ReLU activation. Overall, the GNN provides a differentiable mechanism for aggregating and propagating information across agents. It also accommodates variable network sizes while maintaining permutation invariance with respect to AP and UE ordering. The number of message-passing rounds determines each agent’s spatial communication horizon: one round informs direct neighbors (agents with the same AP or UE), while multiple rounds enable long-range coordination across the network.

Our experiments revealed that $2$ rounds of message passing provides the best overall performance. Beyond two rounds, we observe diminishing performance gains. Additional propagation layers broaden the receptive field but also introduce two drawbacks: (i) information dilution and over-smoothing, where node embeddings become less distinguishable and gradients harder to optimize, and (ii) increased computational and communication overhead, which is undesirable for latency-sensitive RAN control. Conversely, using only one round restricts coordination to very local neighborhoods, preventing agents from capturing multi-hop interference patterns and global network load conditions.

In addition to spatial dependencies, channel quality exhibits temporal structure due to UE mobility. We model this variation via short histories of local channel magnitudes. Each agent processes this history using a Gated Recurrent Unit (GRU), which summarizes temporal dynamics into a compact embedding. The GRU output replaces raw channel samples as input to the GNN, allowing the model to jointly capture temporal evolution (via GRU) and spatial coupling (via GNN). 

\subsection{Constrained RL with Local Reward and Cost} \label{subsec:reward_cost}

The APS objective is to minimize power usage by deactivating APs while ensuring that all UEs are provided with their required SE value. These objectives are inherently conflicting and have distinct natures: SE satisfaction represents a hard constraint, whereas power reduction is an optimization goal. Collapsing them into a single scalar reward would obscure their roles and impede learning the optimized behavior. Therefore, we employ a \emph{constrained reinforcement learning} formulation, where AP activation contributes to the reward, and SE violations are treated as a separate cost. Maximizing the reward while keeping the cost below a prescribed threshold through a constrained reinforcement learning framework naturally aligns with the intrinsic structure and objectives of the APS problem.

In Section~\ref{sect:eval}, we evaluate this constrained RL framework and compare it against an unconstrained RL baseline.

Defining global reward and cost values across all agents leads to credit assignment issues~\cite{foerster2018counterfactual}, especially in large-scale scenarios. When every agent receives identical feedback signals, it becomes difficult to relate individual actions to their effects on performance. To mitigate this, we define local reward and cost values. For agent $i$ controlling AP-UE pair $(m,k)$, we define
\begin{align}
    r_i^n &= - \operatorname{sign}\left[\sum_{k=1}^{K}\alpha_{m,k}^n\right] \\
    c_i^n &= \kappa - \frac{1}{T} \sum_{t\in[T_n]} \mathrm{SE}_k(t)
\end{align}
where $r_i^n$ penalizes AP activation and $c_i^n$ measures the SE shortfall for UE $k$. 

\begin{figure}[t] 
\includegraphics[width=0.9\linewidth]{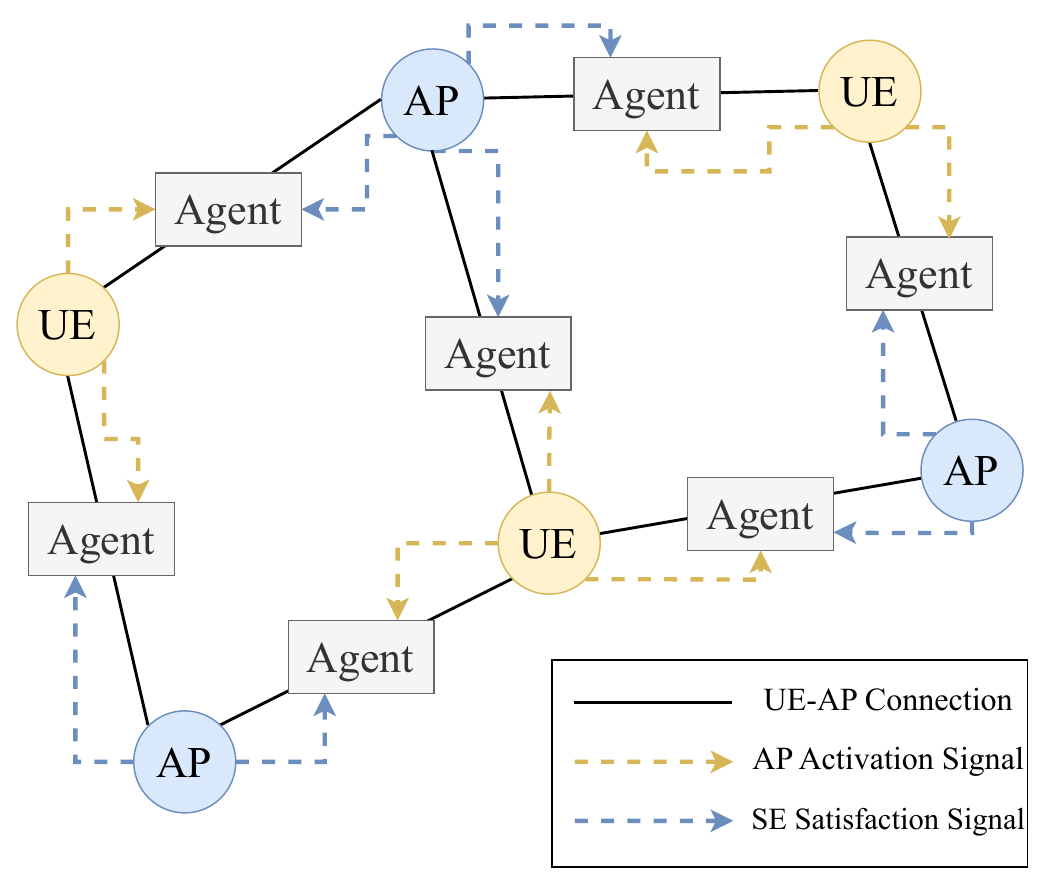}
\caption{Local reward structure for each agent. Each agent associated with an AP-UE pair receives a local reward composed of two signals: the AP activation signal, and the UE SE satisfaction signal.}
\label{fig:reward}
\end{figure}

Cooperation naturally emerges from this structure. All agents linked to the same AP share the same reward, aligning their incentives to avoid redundant activations that waste power. Similarly, agents associated with the same UE share the same cost, discouraging actions that degrade SE. Consequently, coordination arises implicitly from overlapping reward-cost dependencies rather than a shared global feedback. This mechanism, illustrated in Figure~\ref{fig:reward}, enables scalable cooperation through local reinforcement signals.

% Following a fully cooperative MARL design, we assign identical reward and cost signals to all agents. This avoids introducing artificial competition that could destabilize learning or reduce overall system efficiency. At step $n$, the per-agent reward and cost are defined as
% \begin{align}
%     r^n &= -\frac{1}{M} \sum_{m \in [M]} \operatorname{sign}\!\left(\sum_{k \in [K]} \alpha_{m,k}^n\right), \\
%     c^n &= \kappa - \frac{1}{T \times K} \sum_{t \in [T_n]} \sum_{k \in [K]} \mathrm{SE}_k(t).
% \end{align}

% Maintaining the cost below zero enforces that UEs collectively achieve an average SE of at least $\kappa$. The reward structure further incentivizes AP deactivation: when an AP serves no UEs (i.e., all $\alpha_{m,k}^n = 0$), it is switched off, thereby increasing the reward and promoting energy-efficient operation.

\section{APS-GNN Training} \label{sec:training}

\begin{figure*}[t] 
\includegraphics[width=\linewidth]{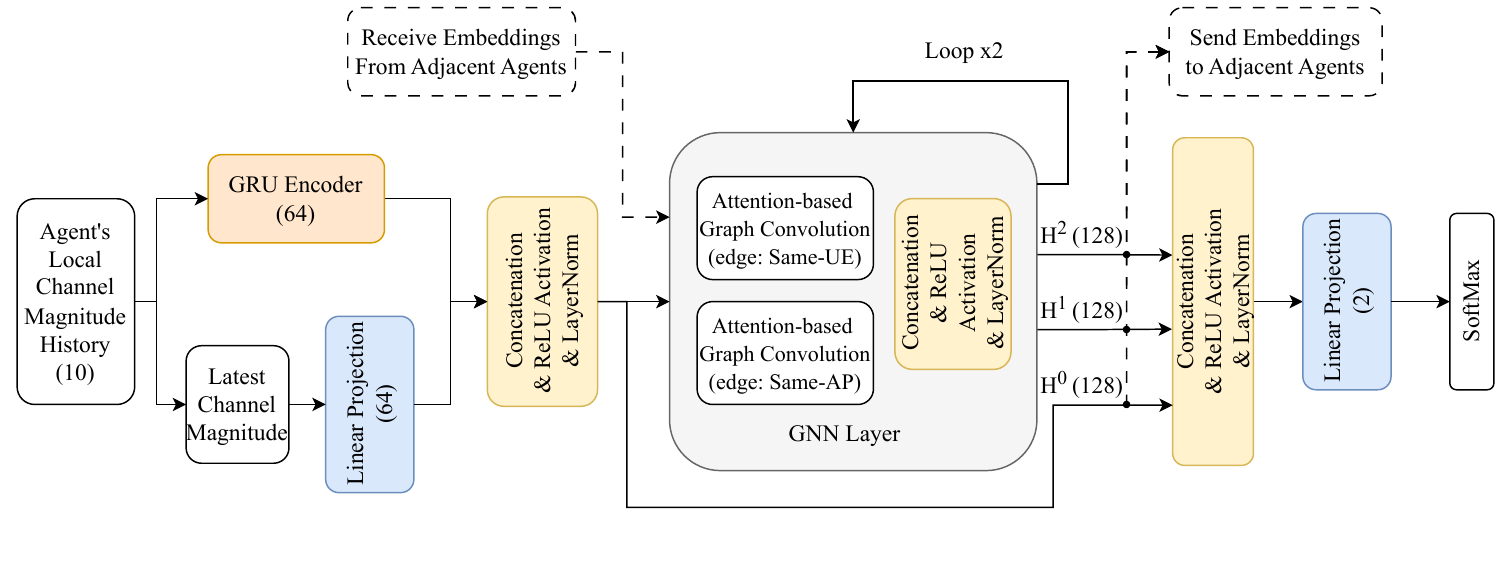}
\caption{Architecture of APS-GNN. Each agent encodes its local temporal channel history using a GRU, which is concatenated with the most recent channel magnitude and projected into a $64$-dimensional embedding. The resulting node features participate in two rounds of message passing using Attention-based Graph Convolution layers over same-UE and same-AP edges, enabling spatial coordination among neighboring agents. After concatenation, ReLU, and LayerNorm operations, the final $384$-dimensional embedding is linearly projected to a $2$-dimensional output representing the binary AP-UE connection decision.}
\label{fig:gnn}
\end{figure*}

\subsection{MAPPO-Lagrangian Training}
We train APS-GNN using a policy-gradient multi-agent reinforcement learning (MARL) approach. Policy-gradient methods update action probabilities through smooth, incremental adjustments rather than the abrupt maximization steps typical of value-based methods. These continuous updates yield steadier learning dynamics and more reliable convergence in high-dimensional, multi-agent settings. 

We adopt an \emph{actor-critic} framework in which the critic provides a learned value estimate that serves as a baseline for the policy-gradient updates, substantially reducing variance compared to vanilla policy-gradient methods\cite{konda1999actor}. To further stabilize training, we employ \emph{parameter sharing}: all agents use a common actor and critic network while operating on agent-specific observations and rewards. This shared parameterization reduces sample complexity, promotes behavioral consistency across agents, and allows experience gathered by one agent to benefit the entire agent population. As the number of possible AP-UE pairs increases, this shared-policy design significantly enhances scalability and data efficiency.

\textbf{Actor-Critic Architecture.}
As described in Section~\ref{sec:design}, each actor receives a short history of its local channel magnitudes, which is processed by a GRU to extract temporal features. The resulting hidden state is then passed to a 2-layer GNN that aggregates information from neighboring agents. Each GNN node produces a two-dimensional output representing the activation probability of its AP-UE pair, from which the action is sampled during training. Figure~\ref{fig:gnn} illustrates this architecture, where Attention-based Graph Convolution refers to the message passing and aggregation operations implemented in~\cite{parlier2024learning} for power control in CFmMIMO and first introduced in~\cite{shi2020masked}. This layer uses a transformer-like permutation-invariant attention mechanism to flexibly and adaptively weight contributions from different neighbors rather than treating all neighbors equally.

At each time step, the environment provides feedback composed of a reward (reflecting activated AP count) and a cost (reflecting SE violations). To handle this constrained setting, we train \emph{two separate critics}, one estimating the cumulative reward and another estimating the cumulative cost. Both critics share the same RNN-GNN backbone as the actor, enabling distributed value estimation where each agent’s critic processes only the messages within its communication horizon.

\textbf{Lagrangian Constraint Handling.}
The actor’s policy is optimized to increase expected reward while discouraging actions that incur excessive cost. We achieve this through a Lagrangian formulation that combines reward and cost advantages. Given the reward advantage $A_r$ and cost advantage $A_c$, the Lagrangian advantage is
\begin{equation}
    A_L = A_r - \lambda \, A_c ,
\end{equation}
where $\lambda$ is a non-negative Lagrange multiplier that adaptively penalizes constraint violations. The multiplier is updated according to the cost:
\begin{equation}
    \lambda \leftarrow \left[\, \lambda +  \frac{\alpha_\lambda}{K} \sum_{k\in[K]} c_k \,\right]_+ ,
\end{equation}
where $\alpha_\lambda$ is the learning rate and $[\,\cdot\,]_+$ denotes projection onto the non-negative reals. This dynamic adjustment increases $\lambda$ when SE constraints are violated and decreases it when they are met, steering the policy toward the desired operation~\cite{zhang2024scalable}.

\textbf{Stable Policy Updates via PPO.}
To prevent large, destabilizing policy shifts, we adopt the Proximal Policy Optimization (PPO) clipped surrogate objective~\cite{schulman2017proximal, yu2022surprising}. Letting $\rho(\theta)$ denote the likelihood ratio between the new and old policies, the actor objective is
\begin{equation}
    \mathcal{L}_{\text{actor}}
    = \mathbb{E}\left[
        \min\!\left(
            \rho(\theta)\,A_{L},\;
            \text{clip}\!\left(\rho(\theta), 1-\epsilon, 1+\epsilon\right) A_L
        \right)
    \right],
\end{equation}
where $\epsilon$ bounds the permissible policy update magnitude. This clipped surrogate constrains the effective step size within a trust region, ensuring smooth and stable training even under rapidly fluctuating interference patterns or channel realizations.

The complete MAPPO-Lagrangian training procedure is summarized in Algorithm~\ref{alg:mappol}.

\begin{algorithm}[t]
\caption{MAPPO-Lagrangian Training of APS-GNN}
\label{alg:mappol}
\begin{algorithmic}[1]
\Require Learning rates $\alpha_\theta,\alpha_v,\alpha_\lambda$; PPO clip $\varepsilon$
\State \textbf{Initialize:} actor parameters $\theta$; reward-critic params $\phi_r$; cost-critic params $\phi_c$; Lagrange multiplier $\lambda \ge 0$
\For{each training iteration}
  \State \textbf{Collect} $\{s^n,a^n,r^n,c^n\}$ for all agents in a batch:
  \State \hspace{0.9em}Each agent gets a history of channel magnitudes
  \State \hspace{0.9em}\textbf{RNN} encodes temporal features $\rightarrow$ embedding
  \State \hspace{0.9em}\textbf{GNN} propagates messages $\rightarrow$ node features
  \State \hspace{0.9em}\textbf{Actor head} outputs $\pi_\theta(a^n\!\mid\!s^n)$; sample $a^n$
  \State \textbf{Compute value targets}: reward $V_{\phi_r}(s)$, cost $V_{\phi_c}(s)$
  \State Estimate reward advantage $A_r$ and cost advantage $A_c$
  \State \textbf{Form Lagrangian advantage:} $A_L \gets A_r - \lambda \, A_c$
  \State \textbf{PPO ratio:} $\rho(\theta) \gets \dfrac{\pi_\theta(a\mid s)}{\pi_{\theta_{\text{old}}}(a\mid s)}$
  \State \textbf{Actor update (clipped surrogate):}
  \[
  \theta \leftarrow \theta + \alpha_\theta \nabla_\theta\,
  \!\min\big(\rho(\theta)\, A_L,\ \clip(\rho(\theta),1-\varepsilon,1+\varepsilon)\, A_L\big)
  \]
  \State \textbf{Reward-critic update:}\quad
  \[
  \phi_r \leftarrow \phi_r - \alpha_v \nabla_{\phi_r}\,\!\big[\big(V_{\phi_r}(s)-\text{target}_r\big)^2\big]
  \]
  \State \textbf{Cost-critic update:}\quad
  \[
  \phi_c \leftarrow \phi_c - \alpha_v \nabla_{\phi_c}\,\!\big[\big(V_{\phi_c}(s)-\text{target}_c\big)^2\big]
  \]
  \State \textbf{Dual update:}
  \[
  \lambda \leftarrow \big[\, \lambda + \tfrac{\alpha_\lambda}{K} \sum_{k\in[K]} c_k \,\big]_+ ,
  \]
\EndFor
\Ensure Learned actor $\theta$ for AP activation decisions
\end{algorithmic}
\end{algorithm}

\subsection{Supervised Pre-Training via Imitation Learning}
Pure reinforcement learning typically starts from a random policy and improves through trial-and-error interaction, which is highly sample-inefficient-especially in wireless environments where each episode involves costly simulations. To accelerate convergence, we \emph{pre-train} the APS-GNN actor using supervised imitation learning before fine-tuning with RL~\cite{schaal1996learning}. This pre-training initializes the actor with a meaningful baseline policy so that subsequent RL optimization refines decisions rather than starting from uninformed exploration. This warm start anchors the policy in a plausible operational region, improving early-stage stability and convergence speed. The approach follows evidence from recent studies~\cite{lee2023supervised} showing that supervised pre-training enhances sample efficiency and in-context learning.

To build the supervised dataset, we generate labels using the \emph{$k$-strongest} heuristic: for each simulated channel realization, the $k$ access points with the strongest links to each UE are selected. This produces labeled actions for all agents under corresponding channel states. Repeating this process across many realizations yields a rich dataset of state-action pairs that encode effective AP-UE association patterns.

The actor network is then trained to imitate this heuristic policy by minimizing the cross-entropy loss between its predicted action probabilities and the heuristic labels. Subsequent reinforcement learning fine-tunes the policy to optimize energy efficiency and SE satisfaction beyond what the heuristic alone can achieve. In this way, supervised pre-training and reinforcement learning act synergistically: the former supplies an informed initialization, while the latter enables adaptive improvement through continuous interaction with the environment.

\subsection{Comparative Baseline: Centralized Multi-agent RL} \label{sec:matlr}
To assess the performance of APS-GNN where agents have a limited view of the system and take actions in parallel, we compare it with a centralized Multi-agent RL algorithm. For this purpose, we extend Multi-Agent Transformer (MAT)~\cite{wen2022multi}, a state-of-the-art unconstrained MARL algorithm, into a constrained RL variant compatible with our setting and call it \textbf{MAT-Lagr}. MAT treats MARL as a sequence-modeling problem in which a transformer jointly encodes all agent embeddings in the encoder part and generates their actions in an autoregressive manner in the decoder part (see~\cite{wen2022multi} for full details). We keep these foundations in MAT-Lagr and add a second encoder dedicated to cost-value estimation, operating in parallel to the original reward-value encoder. Also, the resulting cost embeddings are fused with the original reward embeddings through a fully-connected linear layer before being fed into the transformer decoder, which then produces the autoregressive action sequence.

\textbf{Similarities with APS-GNN.} We train MAT-Lagr using the same MAPPO-Lagrangian method and the same local reward and cost values used for APS-GNN. Also, we decompose the action space at the granularity of AP-UE pairs, each determining whether its corresponding link is activated, as in APS-GNN. By keeping the training procedure identical, we ensure that any performance disparities stem from architectural factors, centralized attention, and sequential decision making, rather than from the training algorithm. 

Similar to APS-GNN, MAT-Lagr employs extensive parameter sharing across agents: the same encoders and decoder are applied uniformly to all agent embeddings along the sequence dimension, with no agent-specific weights in the attention blocks, feed-forward layers, or output heads. 

\textbf{Differences with APS-GNN.} The main difference between MAT-Lagr and APS-GNN lies in information sharing and decision generation. APS-GNN employs structured message passing along AP-shared and UE-shared edges, reflecting the natural sparsity and topology of the CFmMIMO system. In contrast, MAT-Lagr relies on a centralized transformer attention mechanism in which every AP-UE agent attends to every other, creating a fully dense communication pattern. 

The action-generation process also differs fundamentally. MAT-Lagr produces actions sequentially in an autoregressive manner, with each AP-UE decision conditioned on previously generated outputs. While expressive, this introduces a strict computational bottleneck: actions cannot be computed in parallel. APS-GNN, in contrast, produces all decisions simultaneously, leveraging its structured encoding to provide low-latency inference aligned with the synchronous timescales required for CFmMIMO AP selection.

\section{Evaluation} \label{sect:eval}

\subsection{Experiment Setup}
\begin{figure}[t]
    \centering
    \includegraphics[width=0.9\linewidth]{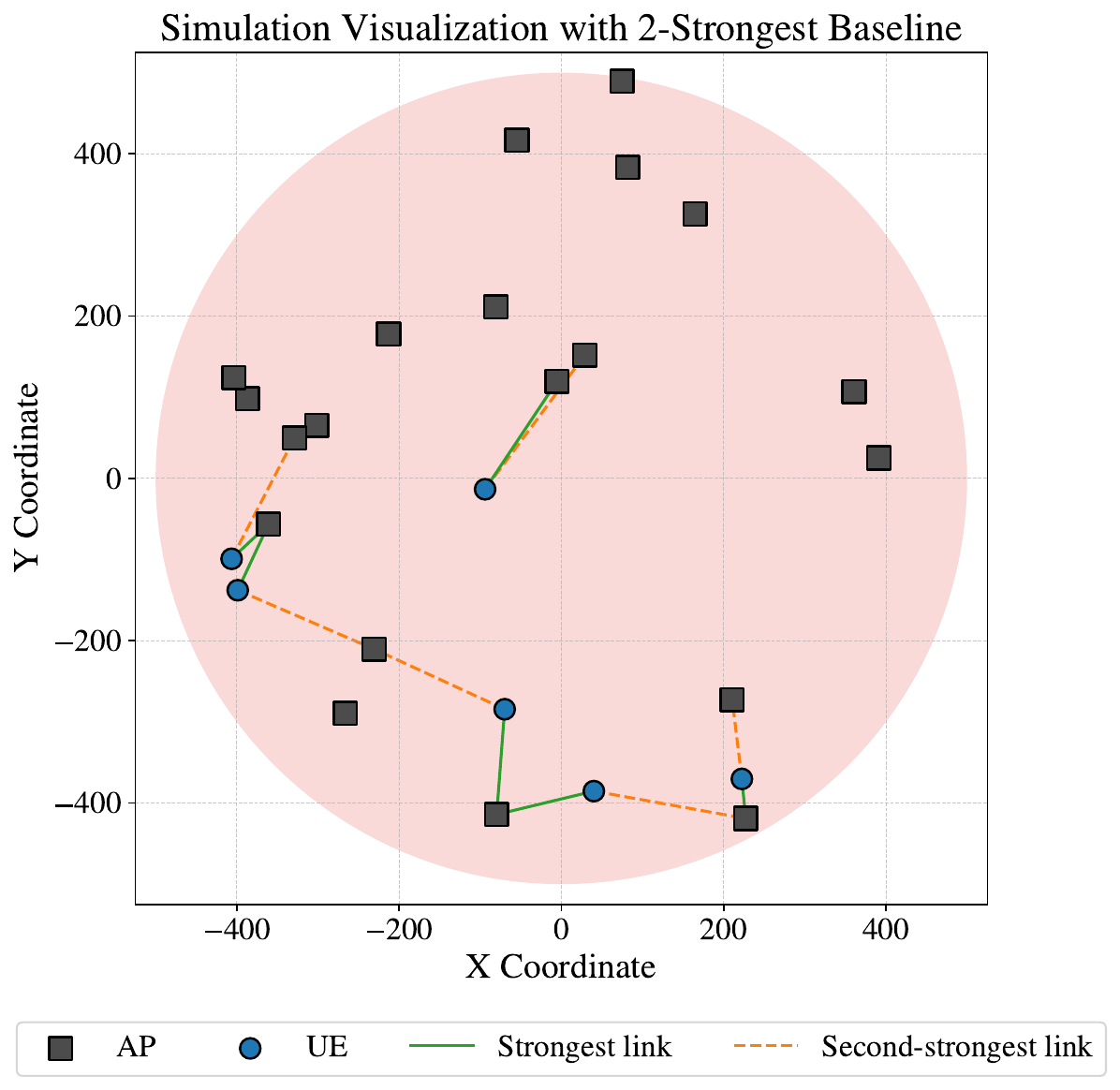}
    \caption{Spatial distribution of APs and UEs within the service area. The plot illustrates UE and AP locations, along with the strongest and second-strongest AP-UE associations.}
    \label{fig:simu_vis}
\end{figure}

To train and evaluate APS-GNN, we developed a CFmMIMO APS simulator that follows the system model described in Section~\ref{sec:sys_model}, aligned with the literature~\cite{9569416, parlier2024learning}. At the beginning of simulation, APs and UEs are placed uniformly at random within the service area, reflecting a spatially unbiased deployment. Figure~\ref{fig:simu_vis} presents an example of such a randomly generated simulation scenario.

UE mobility follows a Random Waypoint mobility model. Each UE selects an initial movement direction uniformly at random, and its speed is sampled from an exponential distribution. The mean of the distribution depends on the mobility class: pedestrian UEs have a low mean speed, whereas vehicular UEs have a significantly higher mean speed. UE positions are updated at each simulation time step based on these mobility parameters. UEs may intermittently pause upon reaching intermediate waypoints. During a pause, the UE remains stationary for a random duration before selecting a new movement direction and speed. This pause behavior captures short-term stops in mobility patterns, such as pedestrians halting or vehicles slowing at intersections, and introduces additional variability into the temporal dynamics of UE locations.

At every time step, the simulator computes the LoS channel coefficients between all AP-UE pairs. The channel gain is calculated as a function of the three-dimensional distance between the AP and UE antennas. Given these channel realizations, the APS mechanism determines which AP-UE links are established, thereby which APs are activated. Power allocation is then performed using MRT on activated APs. The simulation parameters are detailed in Table~\ref{table:simul_params}.

\begin{table}[!h]
    \caption{Simulator parameters}
    \centering
        \begin{tabularx}{0.34\textwidth}[t]{@{}l@{\quad}X@{}}
            \toprule
            \textbf{Channel Condition} & LoS \\
            \textbf{Environment} & Urban \\ 
            \textbf{Precoding Scheme} & MRT \\
            \textbf{Area Radius} & $500$ m \\
            \textbf{Carrier Frequency} & $9$ GHz \\
            \textbf{AP Antenna Height} & $10$m \\
            \textbf{UE Antenna Height} & $1.5$m \\
            \textbf{Pedestrian Speed} & $1$ km/h (Exp.\ Dist.\ Avg.) \\ 
            \textbf{Vehicular Speed} & $35$ km/h (Exp.\ Dist.\ Avg.) \\
            \textbf{APS Window Size} & $50$ ms \\
            \bottomrule
        \end{tabularx}
    \label{table:simul_params}
\end{table}

The resulting channel and power allocation variables are used to compute the received signal power, interference, and the corresponding SE for each UE at that time step. These quantities form the basis of the RL agents' interactions with the environment. The state is constructed from the channel magnitudes of AP-UE pairs. The reward encourages energy efficiency by penalizing the activation of unnecessary APs, while the cost reflects the deviation of the achieved SE from the target requirement. 

To ensure stable and statistically meaningful training, we run eight simulator environments in parallel and collect interaction samples synchronously. Each policy is trained for a total of $500,000$ environment steps, corresponding to APS time windows. All evaluation results reported in this section are obtained over $100,000$ steps. For every scenario and algorithm, we report the mean and standard deviation to account for stochasticity in mobility and channel realizations.

\subsection{Overall APS-GNN Performance Study} \label{subsec:overal}

We study the performance of APS-GNN in scenarios with different network sizes, SE targets, and mobility models. The primary performance criterion is satisfying the SE requirement while minimizing the number of activated APs, on average. We consider the following baselines.

\textbf{k-Strongest Baseline.}
This heuristic connects each UE to the $k$ APs with the strongest instantaneous channels. For every UE, we compute the magnitudes of the channels to all APs, rank them in descending order, and activate the top $k$ AP-UE links, deactivating all remaining links. This procedure yields an $M \times K$ association matrix constructed purely from instantaneous channel strength, without considering interference coupling, power consumption, or SE requirements.

k-Strongest provides a useful lower-bound reference: it is computationally lightweight, requires no training, and exploits the intuitive idea that stronger channels are likely to contribute more effectively to UE performance. However, it does not adaptively balance AP activation costs, does not account for network-wide interactions such as interference across UEs or overloading of heavily used APs, and cannot exploit temporal correlations in the channel process.

\textbf{PPO-Lagr: Single-agent Constrained RL Baseline.}
This baseline quantifies the benefit of decomposing APS into multiple AP-UE-pair agents in APS-GNN, by instead treating the entire AP-UE association matrix as the action of a single monolithic agent. In this formulation, the agent has complete visibility of the network: its state is formed by concatenating the instantaneous channel magnitudes for all AP-UE pairs into a single large vector. Furthermore, since it is infeasible to define local values for a single agent, the cost and reward values are defined as the average values across the whole network.

Training is performed using a PPO-Lagrangian framework, similar to APS-GNN as described in Section~\ref{sec:training}. The agent maintains separate critics for reward and cost estimation and updates its policy via clipped policy-gradient steps with Lagrangian multipliers enforcing the SE constraint in expectation. By holding the learning algorithm constant across baselines, we isolate the effect of moving from a monolithic single-agent architecture to the structured multi-agent APS-GNN design.

\textbf{MAT-Lagr: Centralized Multi-agent Constrained RL Baseline.} A detailed description of this baseline is provided in Section~\ref{sec:matlr}. To enable a fair comparison, MAT-Lagr is trained using the same reward and cost values, optimization procedures, and hyperparameter settings as APS-GNN. The complete set of RL training parameters used across all methods is summarized in Table~\ref{table:rl-params}.

\begin{table}[!t]
    \caption{RL training parameters}
    \centering
    \begin{tabularx}{0.3\textwidth}[t]{@{}l@{\quad}X@{}}
        \toprule
        \textbf{PPO Update Iterations} & $10$ \\ 
        \textbf{PPO Clip Parameter} & $0.1$ \\ 
        \textbf{Gradient Norm Clipping Threshold} & $1$ \\ 
        \textbf{Learning Rate} & $0.001$ \\ 
        \textbf{Gamma} & $0.01$ \\ 
        \textbf{Lagrangian Update Rate} & $2$ \\
        \bottomrule
    \end{tabularx}
    \label{table:rl-params}
\end{table}

Table~\ref{table:method_state_decision} summarizes the information requirements and decision-making structures of all evaluated APS mechanisms.

\begin{table}[!h]
    \caption{Comparison of different APS mechanisms}
    \centering
    \begin{tabularx}{0.48\textwidth}[t]{llX}
        \toprule
        \textbf{Method} & \textbf{Required State} & \textbf{Decision Making} \\
        \midrule
        k-Strongest & Local channel gains & Local, heuristic rule \\
        PPO-Lagr & Global network state & Centralized, single-agent \\
        MAT-Lagr & Global network state & Centralized, per-agent autoregressive decisions \\
        APS-GNN & Local \& neighbor states & Distributed, fully parallel per-agent decisions \\
        \bottomrule
    \end{tabularx}
    \label{table:method_state_decision}
\end{table}

\begin{figure}[t]
    \centering
    \includegraphics[width=\linewidth]{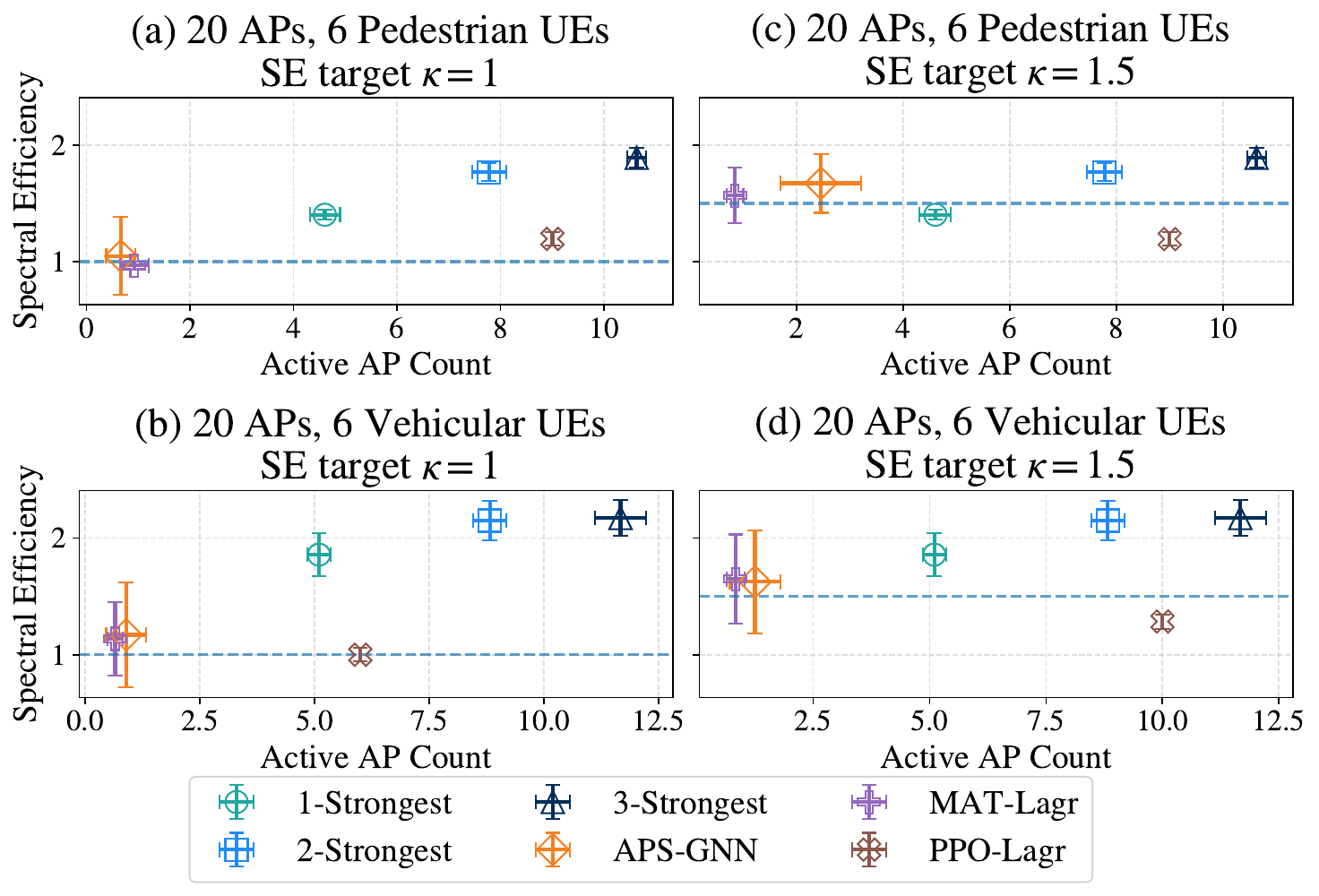}
    \caption{Performance comparison of the algorithms in minimizing the number of activated APs while meeting the target SE across different \emph{medium-scale} scenarios with $20$ APs and $6$ UEs. Error bars indicate the standard deviation.}
    \label{fig:small}
\end{figure}

\begin{figure}[t]
    \centering
    \includegraphics[width=\linewidth]{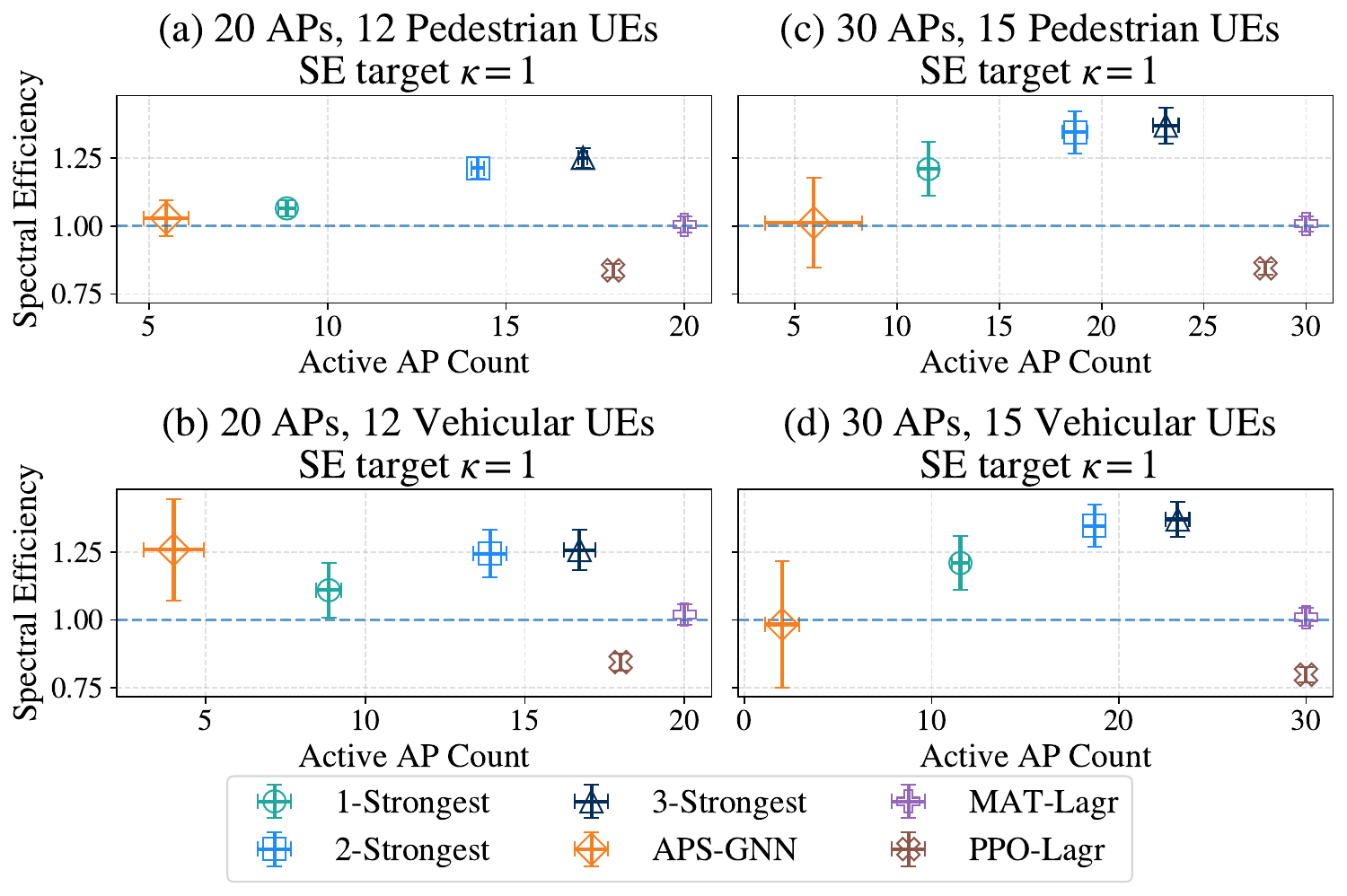}
    \caption{Performance comparison of the algorithms in minimizing the number of activated APs while meeting the target SE across different \emph{large-scale} scenarios. Error bars indicate the standard deviation.}
    \label{fig:large}
\end{figure}

\textbf{Results.} The results are presented in Figure~\ref{fig:small} and ~\ref{fig:large} for medium- and large-scale networks, respectively.

In all medium-scale scenarios in Figure~\ref{fig:small}, APS-GNN and MAT-Lagr deliver the SE target on average while activating the lowest number of APs compared to the baselines. In particular, these two algorithms activate around $70\%$ fewer APs compared to the best-performing other algorithm in these scenarios. APS-GNN achieves nearly the same SE values as MAT-LAgr \emph{and} activates a comparable number of APs. The small performance gap between the two demonstrates that distributed GNN-based decision making can approach centralized MARL performance without relying on global state or sequential action generation. This near-equivalence is particularly meaningful given APS-GNN's fully parallelizable and distributed nature. Nonetheless, MAT-Lagr performs slightly better in providing the target SE with fewer number of activated APs, reflecting the strength of a centralized transformer-based MARL model with global observability. 

In all large-scale scenarios shown in Fig.~\ref{fig:large}, MAT-Lagr consistently activates all APs in order to satisfy the target SE requirements. In contrast, APS-GNN achieves the same SE targets while activating substantially fewer APs, around $50\%$ compared to $1$-Strongest, demonstrating a clear efficiency advantage in large-scale deployments. This performance gap stems from the explicit relational modeling enabled by the GNN, which provides a structured and scalable mechanism for information exchange among agents. As the network size grows, MAT-Lagr must implicitly learn dependencies across an increasingly large set of agents, many of which exhibit weak or irrelevant interactions, leading to inefficient decision-making. By contrast, APS-GNN leverages the underlying graph structure to focus communication and reasoning on meaningful AP-UE relationships, allowing it to scale more effectively and make more selective activation decisions in large networks.

PPO-Lagr is consistently the weakest baseline. The single-agent RL formulation fails to reach the SE targets and tends to activate many APs while yielding low SE. This confirms that collapsing the multi-agent structure into a single-agent setting is ineffective for this task.

Unlike the RL-based methods, the $k$-Strongest baselines do \emph{not} account for the target SE~$\kappa$. They connect each UE to the strongest APs \emph{blindly}, without considering whether this is the best way to meet the required SE target. For instance, under the $1$-Strongest baseline, roughly five APs are activated to serve the six UEs in Figure~\ref{fig:small}. This means that only one AP happens to offer the strongest channel to two UEs, while each of the remaining four UEs is best served by a different AP. Although $1$-Strongest baseline can meet the SE target in three out of the four scenarios in this experiment, APS-GNN and MAT-Lagr show this can be done by activating fewer APs.

\subsection{APS-GNN Ablation Study}
In this section, we further study the impact of different design choices in APS-GNN, particularly, the use of constrained RL, RNN, and pretraining. 

\textbf{Constrained RL Impact.} To assess the role of constrained RL in our design, we construct an \textbf{APS-GNN-Unconstrained} baseline and compare it against APS-GNN. This variant retains all architectural components of APS-GNN but removes the Lagrangian constraint-handling mechanism. Instead, the UE SE and the AP's activation state are combined into a single \emph{engineered} reward. For agent $i$ controlling AP-UE pair $(m,k)$, we define the following reward at step~$n$:
\begin{align}
    r^n = \max\left(\frac{1}{\kappa} \; \frac{\sum_{t \in [T_n]} \mathrm{SE}_k(t)}{T}, 1\right) - \eta \; \operatorname{sign}\!\left(\sum_{k \in [K]} \alpha_{m,k}^n\right)
\end{align}

This reward function incentivizes the policy to increase the average SE until the target $\kappa$ is reached while simultaneously penalizing the AP activation. These objectives inherently conflict: activating more APs improves SE, as illustrated by the $k$-Strongest heuristic in Figure~\ref{fig:small}, but increases power and coordination cost. The hyperparameter~$\eta$ manages this trade-off and must be tuned empirically. However, we argue that the overall performance of APS-GNN-Unconstrained is overly sensitive to this parameter and the optimal value can significantly vary among different scenarios.

\begin{figure}[t]
    \centering
    \includegraphics[width=\linewidth]{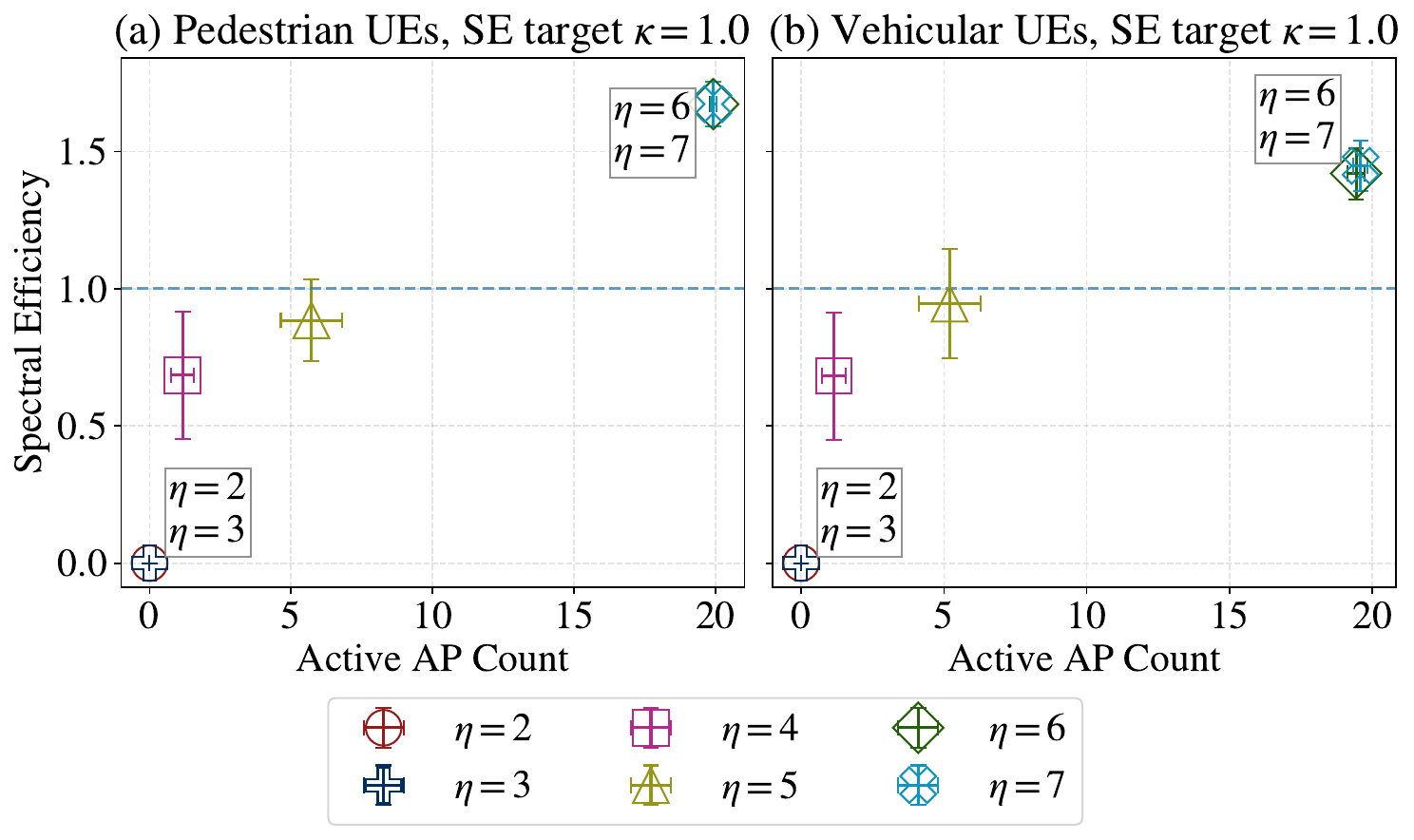}
    \caption{Trade-off between spectral efficiency and activated AP count for APS-GNN-Unconstrained baseline under different parameters~$\eta$. Results are shown for (a) pedestrian and (b) vehicular UE mobility with an SE target of $\kappa = 1.0$. Each marker represents the mean, with error bars indicating standard deviation. Increasing $\eta$ improves spectral efficiency but requires activating more APs, highlighting the inherent objective conflict when collapsing costs and rewards into a single scalar signal.}
    \label{fig:unconstrained_ablation}
\end{figure}

Figure~\ref{fig:unconstrained_ablation} provides the empirical evidence for this argument. Across both pedestrian and vehicular mobility scenarios, different choices of parameter~$\eta$ lead to substantially different operating points. When parameter~$\eta$ is small, the penalty on AP activation dominates the reward, causing the agent to undervalue SE satisfaction. In both pedestrian and vehicular scenarios, the policy aggressively minimizes the number of activated APs-often activating only one or two-at the expense of failing to meet the SE target. Conversely, when~$\eta$ is large, the reward becomes dominated by the SE term, driving the agent to activate many APs in pursuit of higher throughput while largely ignoring the activation penalty. Although intermediate~$\eta$ values may occasionally produce a balanced outcome, the absence of constraint satisfaction guarantees means that performance is highly sensitive to tuning, environment dynamics, and mobility patterns. This variability illustrates that unconstrained RL does not reliably achieve the desired SE target while minimizing AP utilization. In contrast, APS-GNN consistently meets the SE requirement with a lower number of activated APs.

Naturally, one could devise alternative reward functions to balance SE and AP activation. However, any scalarization of these two fundamentally conflicting objectives inevitably requires hand-crafting the trade-off. In unconstrained RL, the designer must manually tune weighting parameters and shape the reward so that the agent neither collapses to activating all APs to satisfy the SE target nor excessively deactivates APs at the expense of violating it. This manual balancing is brittle, problem-specific, and often requires extensive experimentation. In contrast, the constrained RL methodology handles this trade-off in a principled and systematic manner: the constraint is enforced through the Lagrangian mechanism, and the learning process automatically adjusts the dual variables to satisfy the SE requirement without the need for engineered reward terms or painstaking hyperparameter tuning. This distinction highlights the fundamental limitation of the unconstrained approach for the problem formulation we consider.

\begin{figure}[t]
    \centering
    \includegraphics[width=.98\linewidth]{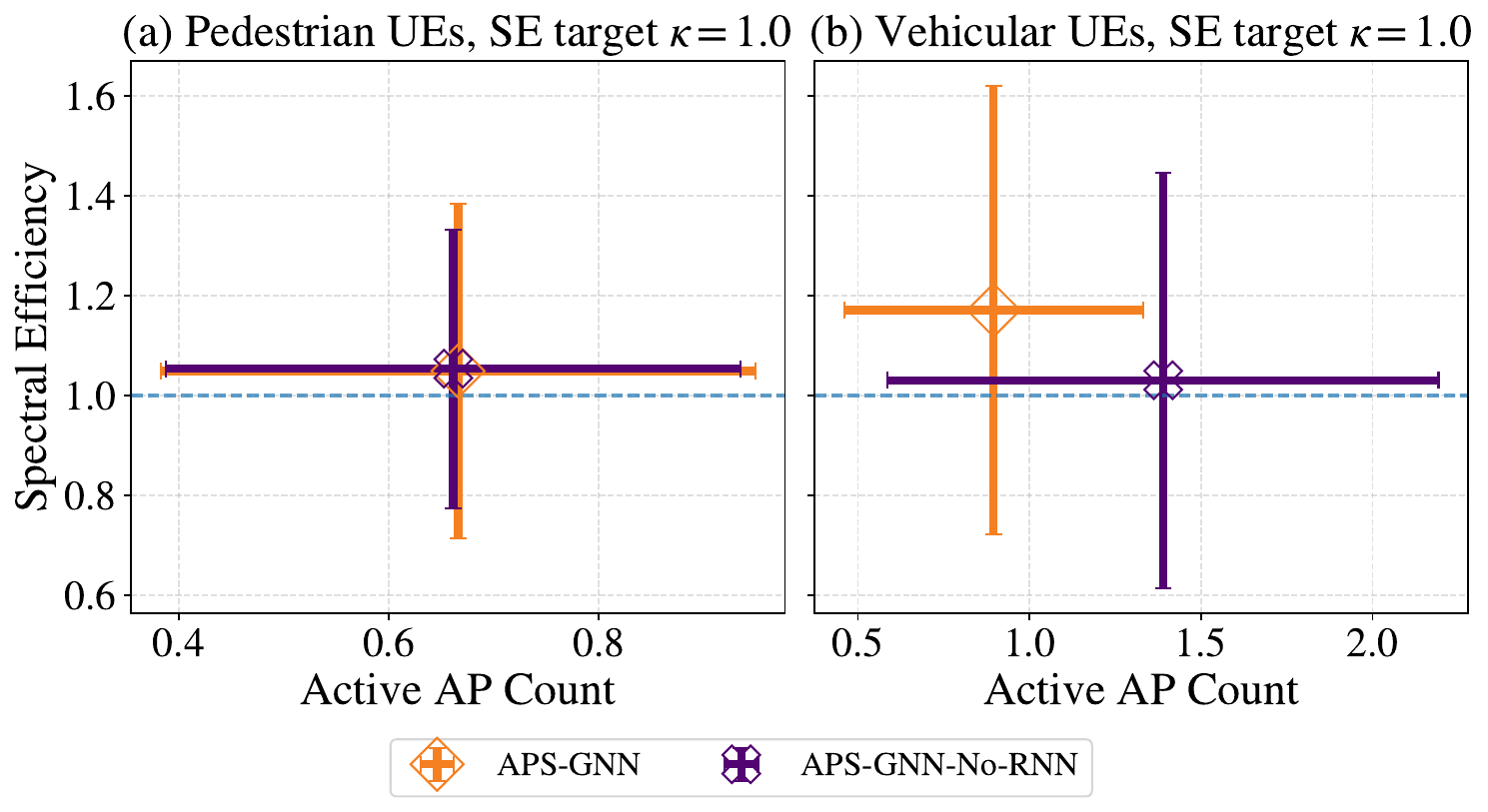}
    \caption{Ablation study of the RNN encoder in two scenarios with $20$ APs and $6$ UEs. Error bars indicate the standard deviation.}
    \label{fig:rnn_ablation}
\end{figure}

\textbf{RNN Impact.} Figure \ref{fig:rnn_ablation} compares the full APS-GNN architecture against APS-GNN-No-RNN, which removes the GRU-based temporal encoder and relies only on instantaneous channel features. In the pedestrian setting, where UE movement is slow and channel variations evolve gradually, the performance gap between APS-GNN and APS-GNN-No-RNN is relatively small. Both models trace comparable trade-offs between activated AP count and achieved SE, indicating that temporal history is less critical when mobility is mild and location changes are limited on average. However, in the vehicular scenario, the difference becomes substantial. Due to the rapid and more pronounced changes in UE positions, instantaneous observations are no longer sufficient. APS-GNN-No-RNN exhibits significantly higher AP activation and less stable SE performance, while the full APS-GNN maintains a much more favorable Pareto frontier. This highlights the importance of temporal modeling under fast mobility: the RNN encoder enables APS-GNN to anticipate short-term channel evolution and avoid unnecessary AP activations.

\begin{figure}[t]
    \centering
    \includegraphics[width=.99\linewidth]{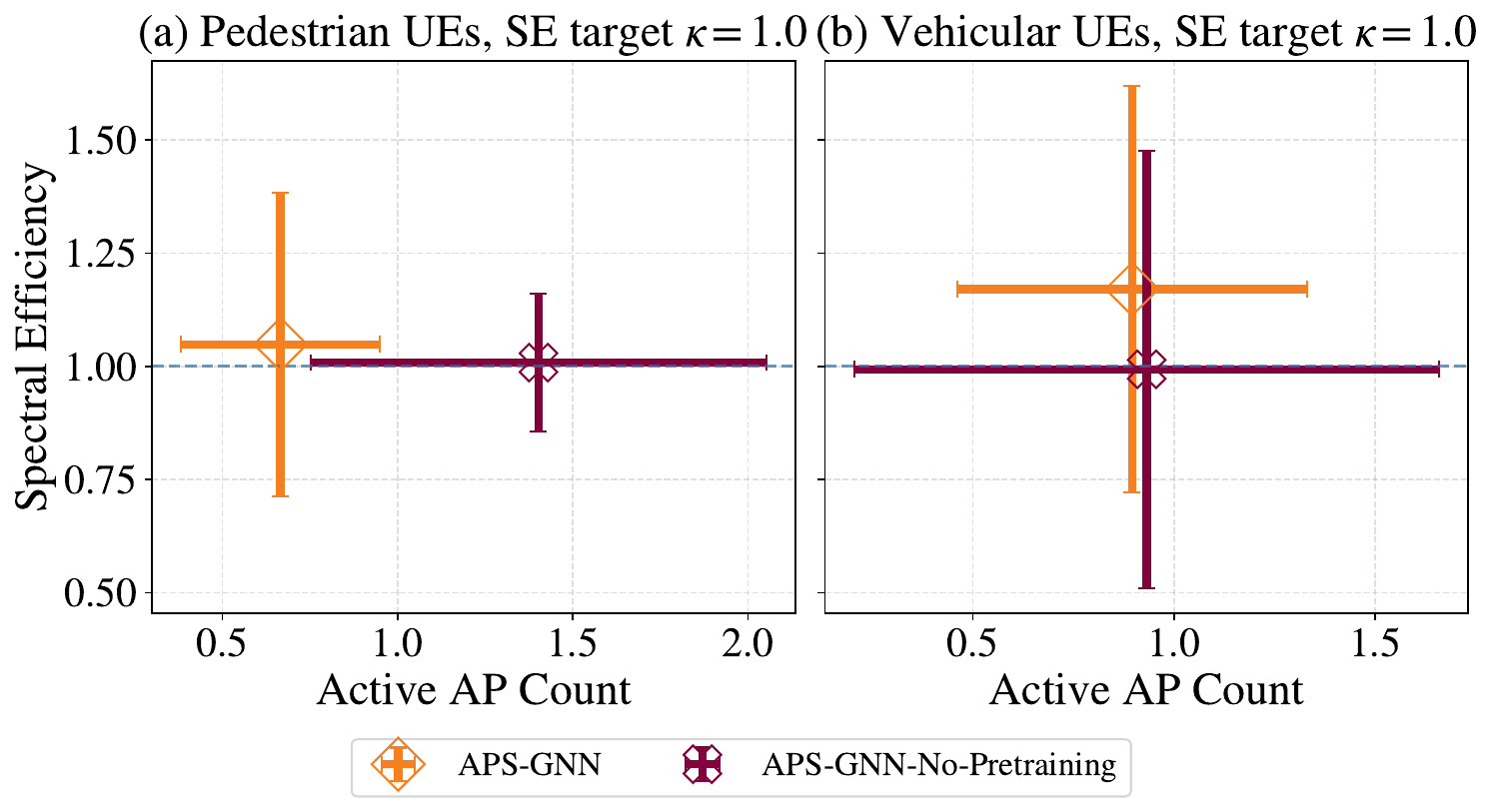}
    \caption{Ablation study of supervised pretraining in two scenarios with $20$ APs and $6$ UEs. Error bars indicate the standard deviation.}
    \label{fig:pret_ablation}
\end{figure}

\textbf{Pre-training Impact.} Figure \ref{fig:pret_ablation} evaluates APS-GNN-No-Pretraining, where the GNN is trained from scratch instead of being initialized through the pretraining stage. In our design, the pretraining dataset is generated using a 4-Strongest heuristic, which constructs association labels by connecting each UE to the four APs with the strongest large-scale fading values. The pretrained model learns to embed AP-UE graph structures and capture useful geometric/channel regularities, which significantly stabilizes subsequent RL fine-tuning. As shown in the figure, removing pretraining shifts the performance curve consistently downwards and to the right.

This indicates that training the GNN from scratch leads to less efficient AP selection policies across both mobility regimes. The gap is especially visible in the pedestrian case, where the pretrained model maintains a substantially better trade-off between activated AP count and achieved SE. Overall, pretraining provides a strong warm start that improves stability and convergence, enabling APS-GNN to operate with fewer active APs while still meeting the SE target.

\subsection{Execution Time}
\begin{figure}[t]
    \centering
    \includegraphics[width=0.95\linewidth]{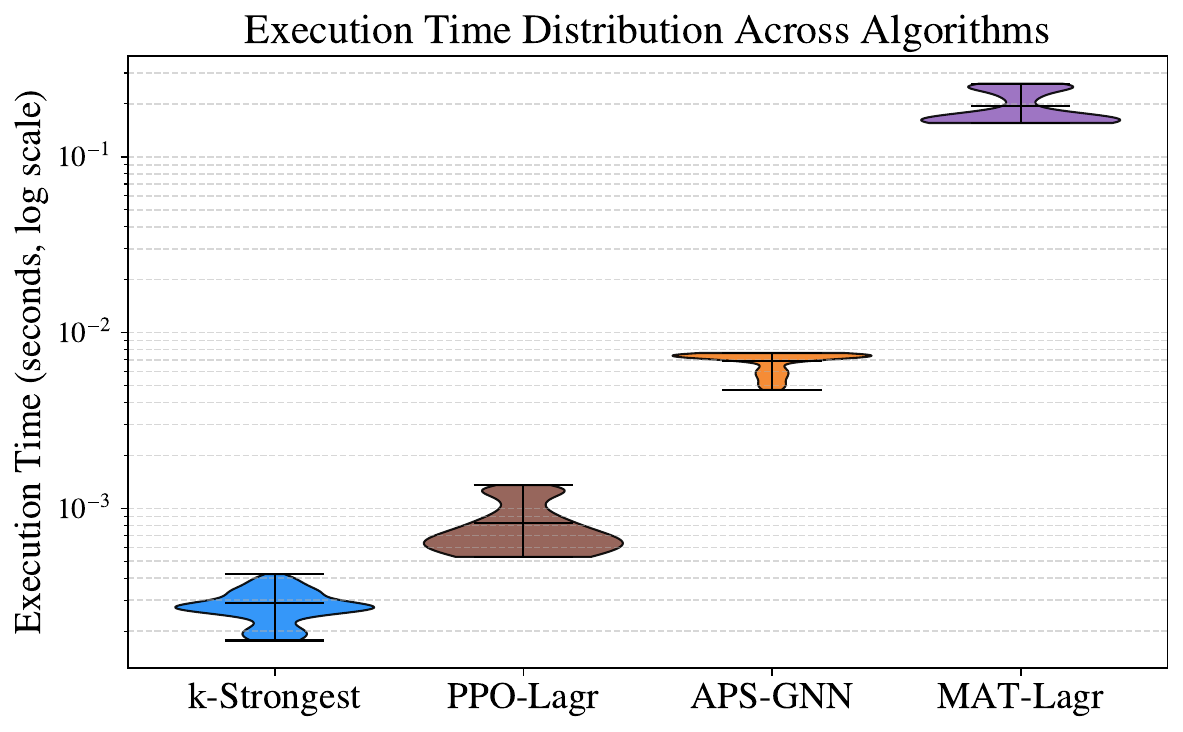}
    \caption{Runtime distribution of all evaluated algorithms in the scenario with $20$ APs and $6$ UEs. Violin plots show the variability and density of inference time on a logarithmic scale.}
    \label{fig:exec_time}
\end{figure}

In this section, we study the execution time of different algorithms evaluated in Section~\ref{subsec:overal}. We run our experiments on a server equipped with an Intel(R) Xeon(R) Silver 4314 CPU (2.40 GHz) and an NVIDIA A40 GPU. While the absolute numbers reported in this section are clearly hardware-dependent, the key takeaway is the relative runtime gap between the different designs, which spans several orders of magnitude. Figure~\ref{fig:exec_time} reports the distribution of per-inference execution time for all four APS mechanisms on a log-scaled axis. 

The non-learning-based k-Strongest heuristic is fastest, with execution times on the order of $0.3$~ms per decision. This is expected, as k-Strongest only requires simple sorting or top-k selection based on channel gains and does not involve any neural network computation. PPO-Lagr remains in the sub-millisecond regime, around $0.7$~ms, since it only performs a single forward pass through a global actor network at inference time. APS-GNN incurs higher inference cost, with runtimes around $5$~ms, reflecting the additional overhead of running a local RNN at each agent together with two rounds of message passing over the AP-UE interaction graph to aggregate neighborhood information before producing actions. Finally, MAT-Lagr is by far the most computationally demanding, with typical execution times around $110$ ms. This is due to both the architectural and algorithmic choices: MAT-Lagr requires running both the actor and critic networks because the architecture relies on value-based embeddings, and the transformer decoder generates actions in an auto-regressive fashion, leading to one decoding step per agent ($120$ agents in this experiment).

Notably, the experiments were conducted on a single machine and the physical channel information propagation delays were not included in the above execution time measurements. The centralized methods such as PPO-Lagr and MAT-Lagr require access to global channel information and therefore would incur additional latency in practice due to the need to collect and aggregate network state at a central controller. In contrast, APS-GNN is designed around local observations and localized message passing, which can significantly reduce the communication overhead and make it more amenable to distributed deployment. 

\section{Conclusion and Future Work}
This paper introduced APS-GNN, a scalable distributed learning framework for the APS problem in CFmMIMO systems. By decomposing APS into per AP-UE connection agents and enabling coordination through graph-based message passing, APS-GNN addresses the scalability and latency limitations of centralized approaches. The proposed constrained RL formulation explicitly captures the conflicting objectives of APS, while local reward and cost signals enable effective credit assignment among agents. Moreover, supervised pre-training provides a principled initialization that anchors the agents to a meaningful operating point, improving exploration efficiency and training stability.

The experimental results consistently demonstrate that APS-GNN delivers the target SE while activating $50-70\%$ fewer APs than baselines. Across medium-scale scenarios, APS-GNN matches the performance of the centralized MAT-Lagr baseline, outperforming heuristic and single-agent baselines. However, in large-scale scenarios, it significantly outperforms MAT-Lagr in deactivating APs, indicating the effectiveness of the structured information sharing among agents. Most notably, APS-GNN achieves these performance gains with low execution times that are an order of magnitude lower than MAT-Lagr.

Several promising directions remain for future work. From a system perspective, APS-GNN can be extended to operate under imperfect or delayed channel state information and to support heterogeneous or time-varying SE requirements across users. From an algorithmic perspective, APS-GNN opens the door to a new class of structure-aware distributed learning algorithms, in which coordination emerges from local message passing rather than centralized optimization. Beyond APS, the proposed framework naturally extends to a wider range of distributed control problems in large-scale wireless networks, including resource scheduling, power control, and joint APS and power control. 
\bibliographystyle{IEEEtran}
\bibliography{mybibliography}

\end{document}